\documentclass[]{spie}  

\usepackage{amsmath,amsfonts,amssymb}
\usepackage{graphicx}
\usepackage[colorlinks=true, allcolors=blue]{hyperref}
\usepackage{xcolor} 
\usepackage{cleveref} 
\usepackage{xurl}

\title{Star Grazing with Alumina Grass: Antireflection coatings in the visible and near-infrared on IPX-Clear Microlenses assisted by Grass-like Alumina}

\author[a,b]{Ishan Rana}
\author[a,b]{Suvrath Mahadevan}
\author[a,b]{Megan Delamer}
\author[a,b]{Ceiwynn Longworth}
\affil[a]{Department of Astronomy \& Astrophysics, Penn State, University Park , State College, USA}
\affil[b]{Center for Exoplanets \& Habitable Worlds,  Penn State, University Park , State College, USA}

\authorinfo{Further author information: (Send correspondence to Ishan Rana)\\Ishan Rana.: E-mail: ikr5146@psu.edu\\}
\begin{document} 
\maketitle

\begin{abstract}
~~~~Two-photon polymerization (2PP) enables the fabrication of high-precision micro-optics with complex freeform geometries, opening up a new parameter space for custom astronomical optics. Among the available resins, the newly developed IPX Clear is particularly well suited for visible applications, offering very high transmission rates across the visible and near-infrared spectrum, low surface roughness, and excellent shape fidelity. However, Fresnel reflections at the air–polymer interface introduce significant optical losses, which are detrimental in low-signal applications such as astronomy, where ever bigger telescopes are being built to collect more photons.
Previous studies have found that grass-like alumina coatings on traditional glass and fused silica lenses improve average transmission from 91.9\% up to 99\% in the 400–900 nm spectral range. In our project, we explore the feasibility of Atomic Layer Deposition (ALD) for applying such a coating to IPX-Clear based micro-optics. In the 400-1700 nm range.
Grass-like alumina based anti-reflective (AR) coatings can be designed to approximate the ideal refractive index for perfect destructive interference. This can be done by generating a gradual refractive index transition from air to the bulk IPX Clear substrate, suppressing surface reflections. While such grass-like coating has been demonstrated on bulk optics, and conformal ALD coatings have been applied to micro-optics built using 2PP. We demonstrate here--- for the first time--- the application of alumina grass coating to microlenses built using 2PP with the new IPX-Clear resin.  In this study we have discussed the challenges and process steps of such applications. We observed Alumina grass coated micro lenses approximately lose 0.3\% of photons due to reflection in the 400-850 nm range. In future work, we plan to test the performance of these microlenses in the entire 400-1700 nm range and explore methods to make coating more resilient, such as by applying an overcoat of SiO\,$_2$.  The combination of these coating with the new and, highly optically transparent IPX-Clear resin opens up a regime of custom designed, highly efficient microlenses for various astronomical applications.

\end{abstract}


\section{INTRODUCTION}
\label{sec:intro}

\subsection{Micro Optics}
\label{microlense}

~~~~Micro-optics are widely used in astronomical instrumentation and telescopes, both ground- and space-based. A common use case is the Shack-Hartmann wavefront sensor\cite{platt_history_2001} --- an array of micro-lenses placed in front of a detector that track the gradient of a wavefront via centroid-shifting. This is a key element of adaptive optics (AO) systems implemented on the Keck 10\,m telescopes \cite{vanDam_characterization_2003,Wizinowich_keck_2020} and other ground-based optical facilities. Similar arrays of micro-lenses have also been incorporated into integral field spectrographs (IFS) like the Spectrographic Areal Unit for Research on Optical Nebulae (SAURON)\cite{Bacon_sauron_2001} on the William Herschel Telescope, or the VIRUS-2\cite{Lee_virus2_2018} instrument on the McDonald 2.7\,m telescope, as a way to feed an enlarged sky image into a multi-object spectrograph. IFS are incorporated into many subfields of astronomy, from galaxy evolution and dynamics to high contrast imaging of young stellar objects and their planetary systems. Some systems, like the Spectro-Polarimetric High-contrast Exoplanet REsearch (SPHERE) on the Very Large Telescope (VLT), contain multiple lenslet arrays to combine extreme adaptive optics with the capabilities of an IFS\cite{Beuzit_sphere_2019}. Micro optics have also been used in fiber double scramblers.\cite{Hunter_scrambling_1992,roy_scrambling_2014,Halverson_efficient_2015} Development work continues to create micro optics out of alternative materials such as silicon\cite{dahal_optimization_2025} and find new ways to integrate lenses on small scales into photonic technology\cite{Arcadi_design_2024}.

\begin{figure} [ht]
   \begin{center}
   \begin{tabular}{c} 
   \includegraphics[width=1.0\textwidth]{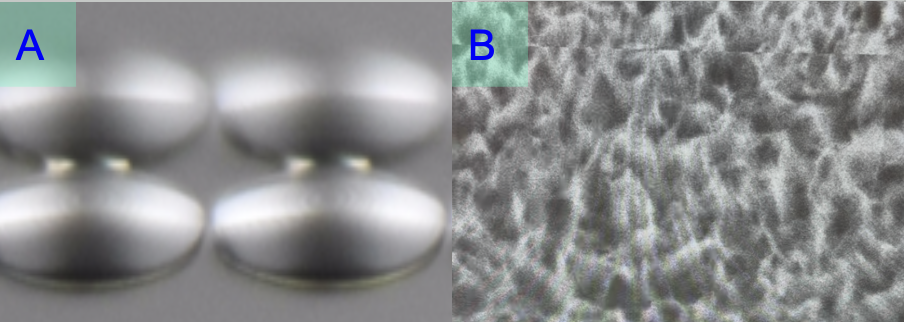}
   \end{tabular}
   \end{center}
   \caption[Lense Diagram]{\label{fig:LenDig} A) Side view of micro lenses being developed for the LFAST project, with diameter = $245 \mathrm{\mu m}$, radius of curvature =$245 \mathrm{\mu m}$ and, height = $19.2\mathrm{\mu m}$. Lens material is IPX clear, and Substrate is Fused Silica. B) SEM image of the Alumina Grass AR coating deposited on the surface of the micro lens. The coating is formed by depositing 35 nm of Alumina bloomed at $90^\circ$\,C for 10 minutes in DI water}
   \end{figure}

Two-photon polymerization (2PP)\cite{goppert_uber_1931,maruo_three_1997} offers the possibility of utilizing complex freeform geometries on lenses smaller than those that can easily be produced through traditional optical manufacturing. The nature of 2PP also allows for precise placement of individual lenses relative to one another or relative to external markers,i.e., it does not necessitate the lenses be on the same underlying rectilinear grid or evenly positioned. The technique also enables scaling from R\&D and prototyping to printing a number of microlenses, which led us to explore the use of this technique for coupling fiber from the Large Fiber Array Spectroscopic Telescope (LFAST)\cite{Bender_LFAST_2022} to its spectroscopic instrumentation. LFAST is a concept for a massively multiplexed array of telescopes,  currently under development,  that combines the light from many small apertures (~0.76\,m) into spectrometers. LFAST will operate in the 400-1700nm range. LFAST can achieve the sensitivity of a much larger telescope at a fraction of the cost of a traditional very large monolithic or segmented mirror. The LFAST fiber combiner uses an array of 19.2\,$\mu$m high, 245\,$\mu$m diameter, and 400\,$\mu$m radius of curvature micro-lenses, \Cref{fig:LenDig}, placed directly over the center of the optical fibers to feed the light into a single square optical fiber.\cite{delamer_pokemon_2025}

\subsection{Anti Reflective Coatings}
\label{ss:ARC}

~~~~Fresnel reflections from the micro-lenses can become a significant source of photon loss, something astronomers are especially sensitive to, given the cost of needing to build ever larger telescopes. To minimize this loss, we developed our micro-lens array using the newly developed and release  IPX Clear resin from Nanoscribe\footnote{Nanoscribe, “Photoresin IPX Clear” \url{https://www.nanoscribe.com/en/products/2pp-printing-materials/photoresin-ipx-clear-for-highly-transparent-microoptics/}.}, which provides high internal transmission ($\sim 99.9\%$) in the visible and near-IR wavelength range. However, at the surface, due to reflections, $\sim$4\% of the photons are lost, which is essentially equivalent to losing 4\% of the telescopes in the telescope array. It is therefore useful to further reduce these losses. The traditional way to address this is to apply a complex multilayer anti-reflective (AR) coatings to the microlens array, but the properties of the IPX-Clear resin precludes using very high temperature coatings.

AR coatings reduce reflection by matching the optical impedance between the incident medium and the substrate. This is done in two main ways \cite{kap}: 

(1) Deposited dielectric thin films: This method uses interference and/or admittance matching to drive the net complex reflection coefficient $r(\lambda,\theta)$ toward zero. In the admittance formulation, the AR condition is \Cref{eq:ra} \cite{TFO}, so the interface presents no mismatch to the incident wave.
\begin{equation}
\label{eq:ra}
Y_0(\theta)-Y_{\mathrm{m}}=0
\end{equation}
This implies that, to make a perfect AR coating using the dielectric layer method,
\begin{equation}
\label{cam}
n_{AR}=\sqrt{n_{\mathrm{substrate}}n_{\mathrm{air}}}
\end{equation}
\begin{equation}
d_{AR} = \frac{\lambda}{4n_{AR}}
\end{equation}
For a perfect AR coating for IPX clear we would need a material with a refractive index of about 1.24 (\Cref{cam}), which does not exist in any bulk material. Also, single layer AR coatings are often narrow-band, but with addition of multiple layers it can be broadened\cite{ristok}. However, we do not wish to subject these lenses to the high temperatures or stresses required or resulting from complex multilayer dielectric AR coatings. Thus, this method is not ideal for LFAST micro-lenses. 

(2) Sub-wavelength surface texturing: To develop AR coatings using this method, one can texture the surface to generate an effective graded-index layer $n_{\mathrm{eff}}(z)$. Such a layer gradually transitions the impedance. AR coatings developed using this method can have effective refractive index in the 1.2 range \cite{kap} (measured using ellipsometer), and such coatings are typically broadband \cite{grad}, making it a promising method for use with IPX clear. 

Near the top surface (the air interface), the effective refractive index ($n_{\mathrm{eff}}$) is closer to that of air, and it gradually increases toward the dense material beneath until it reaches the refractive index of the original surface. To determine the gradient between refractive index at the surface and refractive index at the substrate, previous studies have used gaussian like fits \cite{Fits}. In this study, we use an "Inverse bump function", A bump function is a gaussian like function which has the characteristic bell shape like the gaussian function but it terminates at a set point,

\begin{equation}
n(z) = n_{\mathrm{superstrate}}+(n_{\mathrm{base}}-n_{\mathrm{superstrate}})\sqrt{\frac{\ln{(\frac{d}{z})}}{\ln{(\frac{d}{z})} +\alpha}}
\label{eq:1}
\end{equation}

Here $n$($z$) is the function that gives the effective refractive index at height $z$ above the base of the gradient material. $d$ is the effective height of the gradient material (can be measured using ellipsometer or be a fitting parameter), $n_{\mathrm{base}}$ is the refractive index of the gradient material at its base, $n_{\mathrm{superstrate}}$ is the refractive index of the superstrate (in this case, air), and $\alpha$ is a fitting parameter that affects the width of the inverse bump.

We use the inverse bump function instead of an inverse gaussian function because the effective refractive index has a hard cut-off point at the base and at the air interface, i.e., refractive index there must be equal to that of original ALD Alumina layer and air, respectively. 
 
In this paper, we demonstrate an AR coating solution for 2PP-fabricated micro-optics \Cref{microlense} using gradient refractive index effect of nanoporous amorphous alumina coatings and quantify the resulting improvement in throughput.

\subsubsection{Atomic Layer Deposition}

~~~~Traditional coating methods such as plasma-enhanced chemical vapor deposition \cite{Sput}, 3D printing of moth-eye type coatings \cite{Moth}, and ion-assisted electron beam evaporation \cite{IBE} work well for creating coatings on simple surfaces, but they have limitations when used on complex surfaces such as multiple micro-optics or complex free form micro-optics\cite{ristok}. These techniques are largely directional and, therefore, cannot provide uniform coverage. For micro-optics like LFAST micro-lenses instead require conformal deposition methods like Atomic layer deposition (ALD).

ALD is a thin-film fabrication process that deposits atomic layers on a surface through self-limiting surface reactions\cite{ALDrev}. In each cycle, the substrate is exposed to alternating chemical precursors that react only with surface sites, producing highly uniform and conformal films even on complex, irregular structures. This makes ALD well suited for AR coatings on micro-optics. ALD has been used to deposit various single and multiple dielectric layers on micro-optic surfaces\cite{ristok,surface,smoothsio2}, demonstrating the proof of concept. In this work, we show that ALD techniques can yield high performance and broad-band (400-1700nm) AR coatings on IPX-Clear microlenses, making it suitable for astronomical applications.

\subsubsection{Alumina Grass}
\label{AG}

~~~~Alumina Grass refers to the nano-porous structures that develop on the surface of ALD alumina after exposure to hot deionized (DI) water ($\geq 50^\circ\mathrm{C}$)\cite{kap}. It is well established that alumina films can structurally transform in the presence of moisture and, in particular, during hot-water treatment,\cite{Aby,Vip} giving rise to these characteristic surface structures. Building on this corrosive behavior, some studies \cite{kap,benq,kap21,Isa} intentionally bloomed the resulting morphology to study its use as an AR coating on materials such as fused silica and D263 glass. The results were promising, with an average reflectance of 1\% in the 350–800 nm range in Kauppinen et al.(2017)\cite{kap}, and $\leq0.1\%$ in the 400-700 nm range in Guo et al. (2025)\cite{benq}. We apply this technique to develop and characterize AR coatings to test on the LFAST micro-lenses.

\section{BUILDING THE AR COATING}
\label{sec:process}

\subsection{ALD Chamber}

\begin{figure} [ht]
   \begin{center}
   \begin{tabular}{c} 
   \includegraphics[height=9cm]{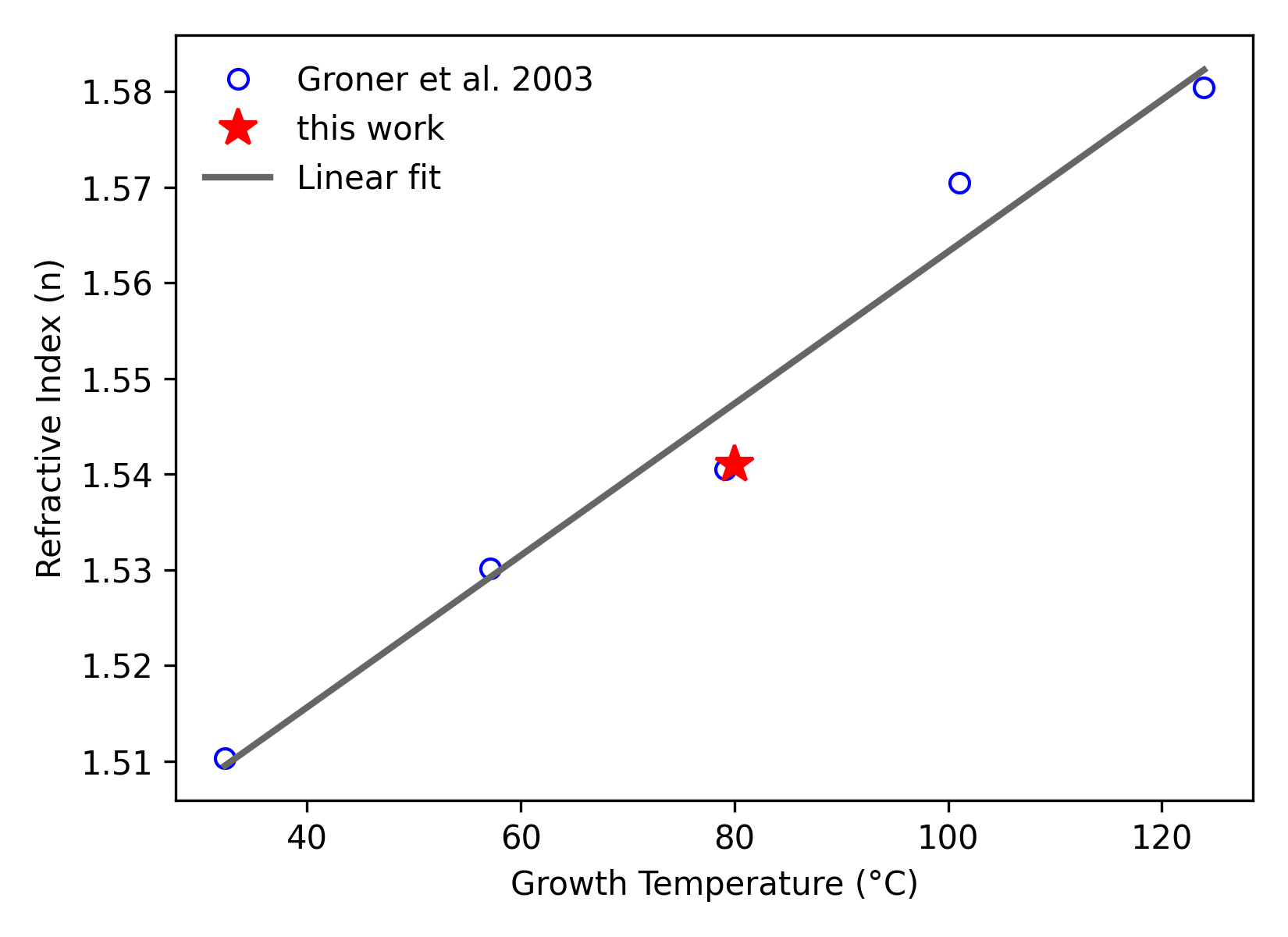}
   \end{tabular}
   \end{center}
   \caption[SEM Grass]{\label{fig:SEMg} Change in refractive index of the alumina (y axis) with change in ALD chamber/Growth temperature (x axis). The data points are from \cite{Groner}. Here, we perform a linear fit to estimate the Alumina refractive index given a deposition temperature. We found the relation to be $n(T) = 1.48 + 7.96\times10^{-4} T$. This study also performed physical measurements using an ellipsometer (Red Star) for alumina deposited at $80^\circ\mathrm{C}$. The results were nearly identical to previous work.\cite{Groner}}
   \end{figure}

~~~~To build the AR coatings, we used the K.J. Lesker ALD 150 LE \footnote{\url{https://www.lesker.com/process-equipment-division/thin-film-systems/ald-150le-atomic-layer-deposition-ald-system.cfm}}
, a thermal ALD instrument at the Penn State Nanofabrication Facility. We deposit amorphous ALD alumina on the micro-lenses using the water–trimethylaluminum (TMA) process \cite{ALDrev} at a chamber temperature of $80^\circ\mathrm{C}$. We choose $80^\circ\mathrm{C}$ because alumina deposited at lower temperatures is less dense and has a lower effective refractive index (1.54). The value at this temperature is close to the refractive index of IPX Clear over our wavelength range of interest (400 nm to 1700 nm), which is desirable when there are no additional layers between the Alumina grass and IPX clear substrate. \Cref{fig:SEMg} shows the change in refractive index with respect to ALD chamber temperature. 

\subsection{Blooming with DI Water}

~~~~After ALD deposition, we treat the surface with deionized (DI) water to promote the growth of the nano-porous structures described in \Cref{AG}. This step transforms the smooth amorphous alumina into a porous, high-surface-area layer with a local packing fraction that varies with height. This resulting structure broadens AR performance and suppresses reflection across a wavelength band wider than a single abrupt interface.

\subsubsection{DI Water Temperature}

\begin{figure} [ht]
   \begin{center}
   \begin{tabular}{c} 
   \includegraphics[height=9cm]{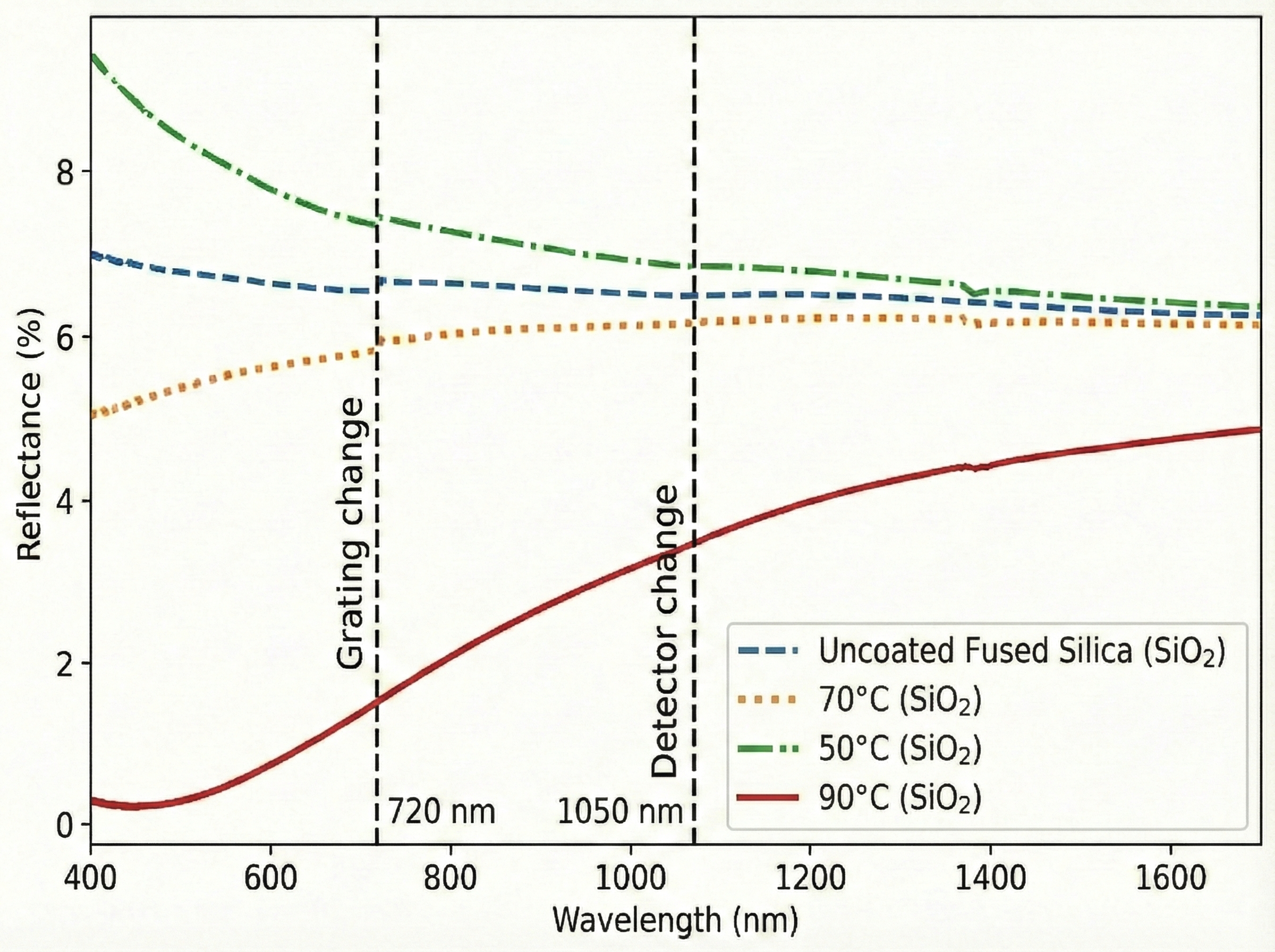}
   \end{tabular}
   \end{center}
   \caption[Temp]{\label{fig:Temp} Reflectance\% vs wavelength relation of an uncoated fused silica slide and 3 double sided ALD Alumna coated fused silica slides bloomed at different temperatures. All slides were coated with 35 nm of Alumina using ALD at a chamber temperature of $80^\circ C$. Each slide was then bloomed in a DI water bath for 10 minutes at $50^\circ C$, $70^\circ C$, and $90^\circ C$ respectively. We then measure the Reflectance vs Wavelength relation of the bloomed slides and the uncoated slide. The slight kink observed at the
   $\sim700nm$ is due to grating change in the Carry 7000.}
   \end{figure}

~~~~DI water temperature can affect the resulting “grass” morphology. In general, higher-temperature blooming produces a stronger anti-reflection effect because it promotes the formation of a more porous nanostructured layer on shorter timescales. A well-developed “grass” layer creates a smoother transition in effective index from air ($n \approx 1$) to dense alumina, which reduces Fresnel mismatch and suppresses reflection across a wider band. For this reason, many prior studies bloom alumina at temperatures $\sim 90\textdegree\mathrm{C}-80\textdegree\mathrm{C}$), where the graded-index morphology forms reliably and yields improved reflectance outcomes\cite{kap,benq} . \Cref{fig:Temp} shows the reflection performance of 3 identical fused silica slides, coated with an identical ALD alumina layer (35 $\mathrm{nm}$ at 80$^\circ\mathrm{C}$) and bloomed at three different temperatures ($50^\circ C$, $70^\circ C$,$90^\circ C$) for a set time (10 minutes).

We observe slides bloomed at high temperatures have better reflectance profiles. We also see that the slide bloomed at $50^\circ C$ did not display any AR behavior. this is most likely due to the fact that the AR coating has not had enough time to  fully bloom, although we note that in previous works \cite{kap}, blooming was observed after 30 minutes of immersion in water at $50^\circ C$. 

\subsubsection{Maintaining Sample Quality}

~~~~For optimal results, it is essential to use high-quality deionized (DI) water (over $1\mathrm{M\Omega}$ as measured in air) for blooming samples. Using lower-quality water, for example, lab distilled water, instead of DI water introduces dissolved ions and particulates that can dry into residues or streaks on the sample surface. This degrades optical performance by reducing transmittance and increases scattering. After blooming, we dry the samples with a nitrogen blow to remove residual droplets and minimize streak formation. The quality of DI water used to bloom samples in this work is summarized in \Cref{tab:water_quality}, which shows the quality of  DI  water measured in the Penn State Nanofabrication facility and the distilled water measured in our lab using The Ultra-meter III\texttrademark{} 9P AHL Titration Kit. 
\footnote{Myron L\textsuperscript{\textregistered} Company, “ULTRAMETER III\texttrademark{} 9P,” product page, \url{https://www.myronl.com/products/accessories-cases/ultrameter-iii-9p/}.}

\begin{table}[t]
\caption{Water quality measurements for Distilled vs Deionized Water.}
\label{tab:water_quality}
\centering
\begin{tabular}{lcc}
\hline
\textbf{Parameter} & \textbf{Distilled Water (Lab)} & \textbf{Deionized Water (NanoFab)} \\
\hline
Resistivity (k$\Omega$) & 436  & 1740 \\
TDS (ppm)               & 1.27 & 0.28 \\
Alkalinity (ppm)        & 1.2  & 0.2 \\
pH                      & 7    & 0.5 \\
ORP                     & 360  & 120 \\
\hline
\end{tabular}
\end{table}

\begin{figure} [ht]
   \begin{center}
   \begin{tabular}{c} 
   \includegraphics[height=9cm]{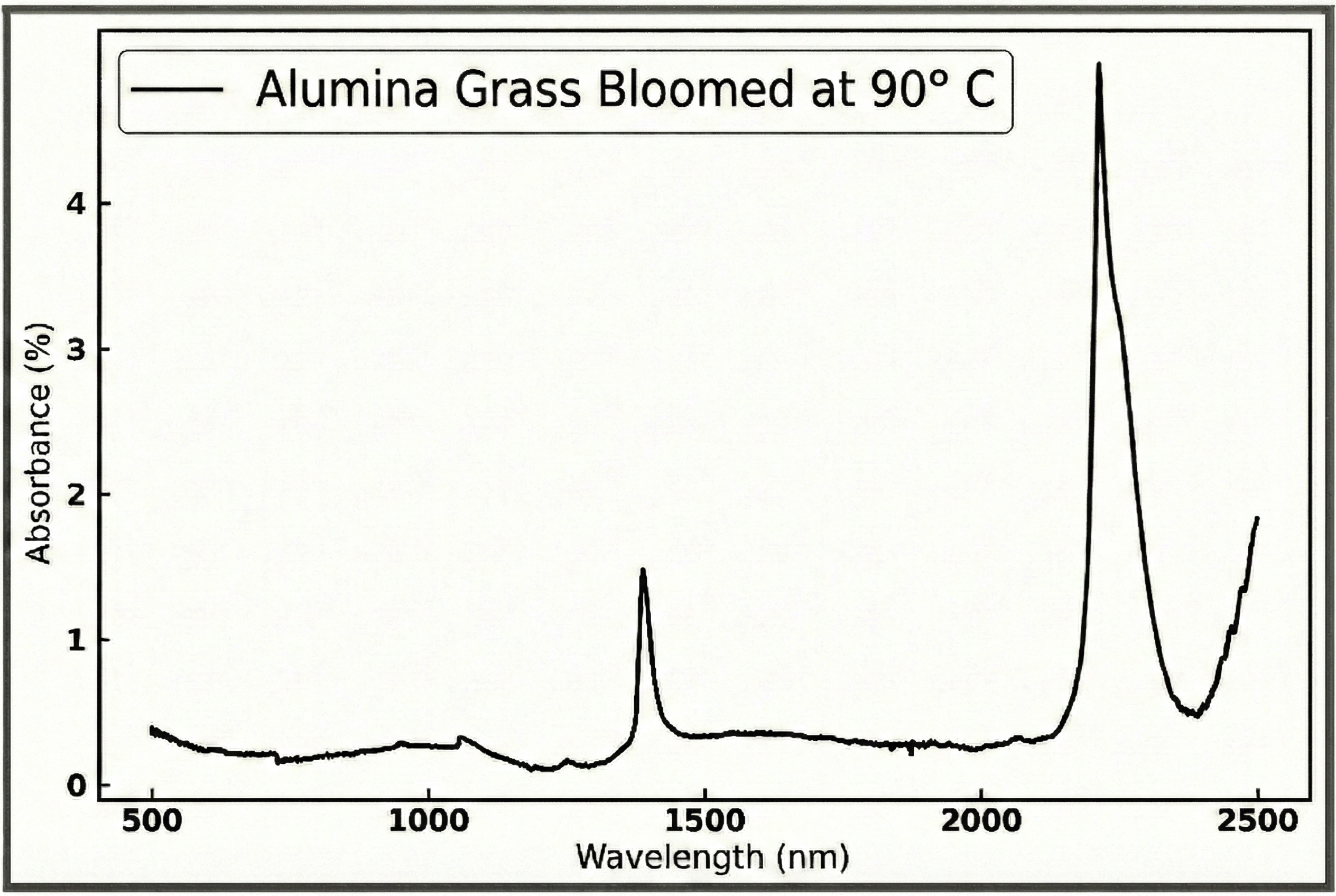}
   \end{tabular}
   \end{center}
   \caption[water]{\label{fig:water} Absorbance\% of a fused silica slide that was initially coated with 35 nm of alumina on both sides and then bloomed at $90^\circ\mathrm{C}$. where $A=1-T-R$. Residual moisture on the AR coatings. We see two absorption features at $\sim1390 nm$ and $\sim2200 nm$, which indicates the presence of moisture. This can be attributed to the porous nature of alumina grass. To avoid this issue we suggest baking the samples after blooming. It is also been observed that blooming the samples in DI water for longer period of time reduces reflectance\cite{Yang}. }
   \end{figure}

After blow-drying the samples with high purity nitrogen gas, some moisture may remain on the sample due to the amorphous nature of the alumina grass. \Cref{fig:water} shows dips in transmission around 1450 nm and 2200 nm, which coincide with well-known near-IR absorption bands associated with O--H species. The remaining moisture can be removed by treating samples in an  oven at $85^\circ C$ \cite{benq}.

\subsection{Simulations}
\label{sims}
\begin{figure} [ht]
   \begin{center}
   \begin{tabular}{c} 
   \includegraphics[height=7cm]{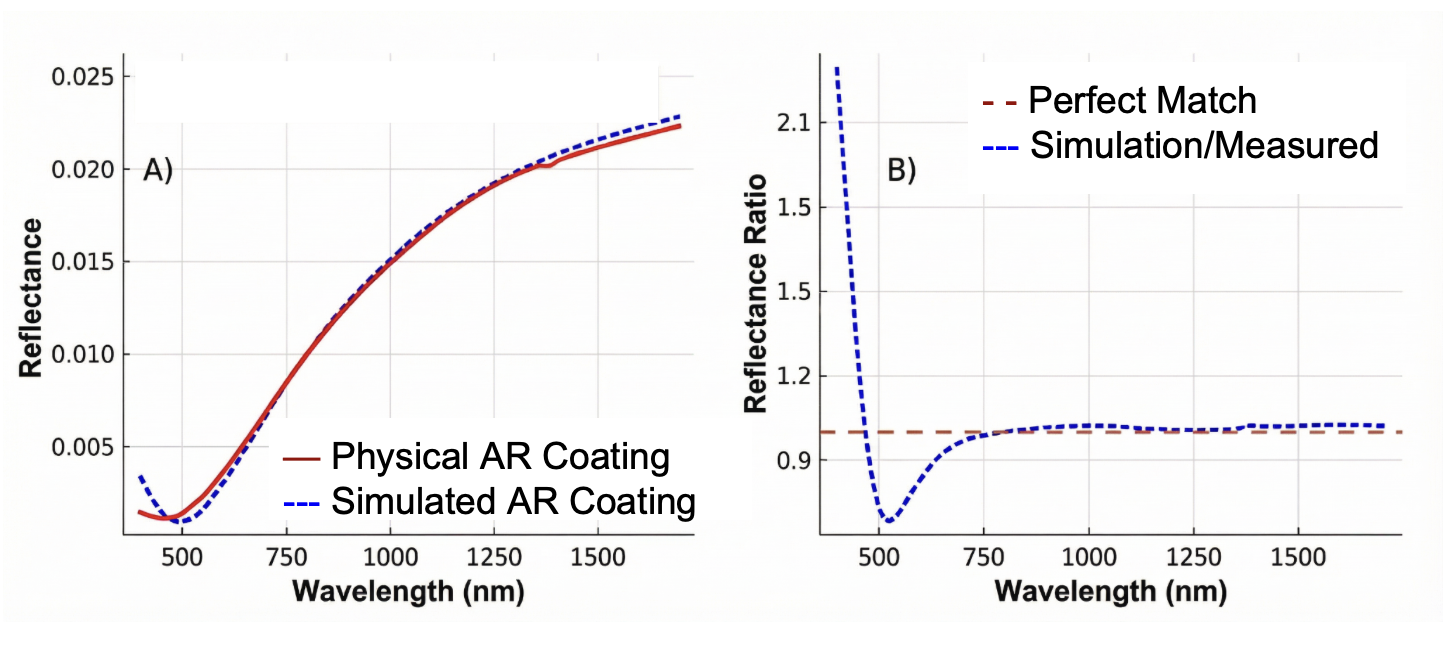}
   \end{tabular}
   \end{center}
   \caption[doubles]{\label{fig:doubles} A) Back reflection separated Measured Reflectance (red) and Simulated Reflectance profile using optimized parameters of the bump function. B) Ratio of Blue over red curve from A). The Ratio is nearly one for IR and red wavelengths, while it deviates from 1 for blue wavelengths, indicating that the optimization yields an  excellent match in the red and less so in the blue. This is likely due to scattering of blue light caused by the irregular structure of alumina grass, which is present in the data, but not accounted for in the simulations.}
   \end{figure}

~~~~Since the Alumina Grass layer optically behaves like a graded-index medium, we can simulate the performance of the coating using the transfer matrix method\cite{GTT}. This is done by assuming a multilayer coating system where all layers have equal thickness (1 nm) \cite{kap} and the refractive index of each layer is based on a unique $n_{\mathrm{eff}}(Z)$ relation using \Cref{eq:1}. This can be found by fitting the relation to the reflectance performance measured by spectrophotometer. \Cref{fig:doubles} shows the agreement of our fit to the measured reflectance profile of a double sided Alumina grass coating bloomed at $90^\circ\mathrm{C}$ and 35 $\mathrm{nm}$ ALD alumina layer. 

We observe the fit agrees well with measured values for red/IR wavelengths average of 1.7\% deviation. In Addition, the calculated height of AR coating was 150 $\mathrm{nm}$, which is within the expected range. However, the fit does not perform well for blue wavelength range due to higher scattering. Other studies have encountered similar issues when fitting gradient profile in the blue wavelength regime \cite{kap} due to scattering,

If the sample is thin, reflectance measurements of samples using a spectrometer will detect some reflections from the rear surface of the sample. Contributions from only the surface of the sample can be calculated using \Cref{eq:2}, if both sides are identical (double side coatings) and assuming there is no internal transmission or scattering losses, i.e., Absorbance $\sim$ 0\cite{Lak}

\begin{equation}
\label{eq:2}
R=\frac{R_\mathrm{measured}}{(2-R_\mathrm{measured})}
\end{equation}

\subsection{Combining Alumina Grass with Dielectric Method} 

\begin{figure} [ht]
   \begin{center}
   \begin{tabular}{c} 
   \includegraphics[height=9cm]{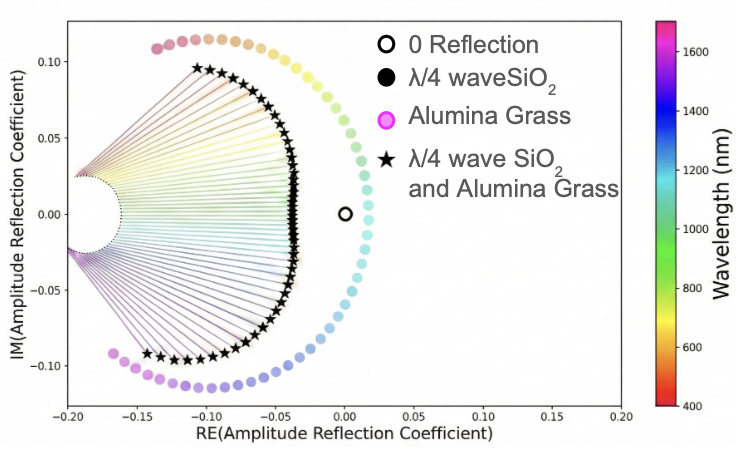}
   \end{tabular}
   \end{center}
   \caption[round]{\label{fig:round} The following is a simulated modified circle diagram. Circle diagrams are typically used to track how the reflectance at a single wavelength changes as each layer is added to a multilayer coating; the final point on the plot gives the total reflectance of the full stack. Here, we apply the same idea across a wavelength range by plotting the terminal points for wavelengths from 400 nm to 1700 nm.
We compare two systems. The first layer is a quater-wave thick coating with refractive index 1.22, which is effective refractive index of our alumina grass samples bloomed at $90^\circ C$ (Initial Alumina coating 35 nm). This layer is A) on a glass substrate of refractive index 1.55 and B) on a stack of 90 nm silica layer over glass substrate. The x- and y-axes represent the complex plane of the amplitude reflection coefficient. For each wavelength, the distance of the terminal point from the origin corresponds to the final reflectance at that wavelength. Points closer to the origin indicate lower reflectance and therefore better coating performance.
We find that adding the silica underlayer flattens the wavelength-dependent curve and, overall, moves the terminal points closer to the origin. This reduces reflectance across a broader wavelength range, although it can increase reflectance at some wavelengths compared to the Alumina Grass alone. }
   \end{figure}

\begin{figure} [ht]
   \begin{center}
   \begin{tabular}{c} 
   \includegraphics[height=9cm]{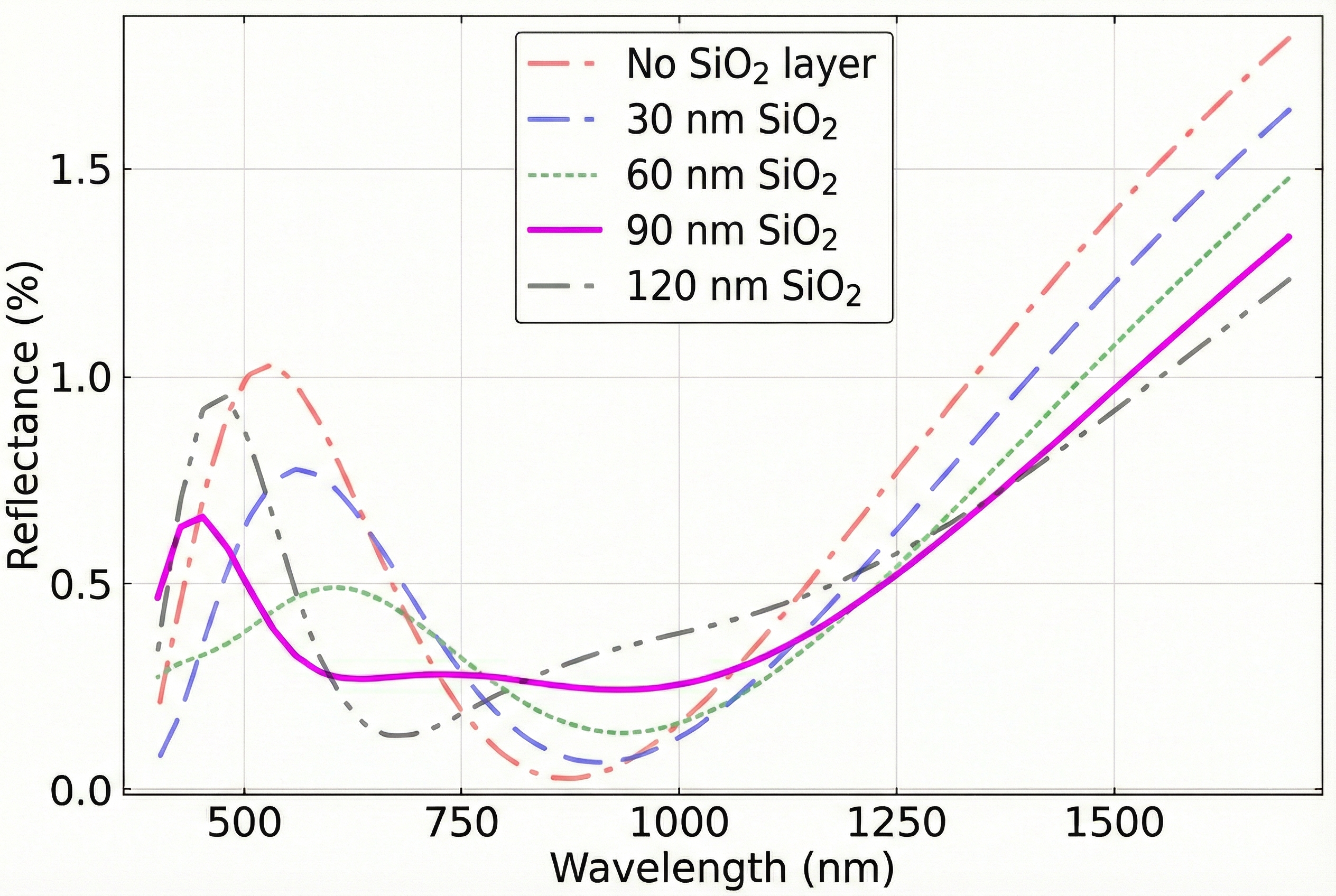}
   \end{tabular}
   \end{center}
   \caption[Silc]{\label{fig:silc} Simulation of Changes in the Reflectance profile of AR coated Soda-lime slides when an undercoat of fused silica is added between the coating and substrate. The addition of the fused silica layer broadens the reflectance profile, increasing reflectance at some points but overall reducing reflectance.}
   \end{figure}

~~~~To further broaden the anti-reflective effect across a wider wavelength range, we can add dielectric coatings ref \Cref{ss:ARC} like silica (SiO$_2$) between the Alumina Grass and Substrate\cite{benq}. \Cref{fig:round} shows simulated results of a circle diagram illustrating how the addition of the underlying silica layer shifts the permittivity curve of the sample. \Cref{fig:silc} shows simulated results of how the thickness of the silica layer affects the reflectance curve of the sample. we observe that a 90 $\mathrm{nm}$ layer gives the best broadband effect as it is close to the quater wave thickness. 
   
\section{Performance of Alumina Grass}

We measure the performance of the AR coating using the Agilent Cary 7000 UV-Vis-NIR ~~~~spectrometer \footnote{\url{https://www.agilent.com/en/product/molecular-spectroscopy/uv-vis-uv-vis-nir-spectroscopy/uv-vis-uv-vis-nir-systems/cary-5000-uv-vis-nir?srsltid=AfmBOoor9cA1r9YrTQQcR7UA9XwXjf6Eokc101nlVS6X085c9SFi0BcI}}
and the Universal Measurement Accessory (UMA)\footnote{\url{https://www.agilent.com/en/product/molecular-spectroscopy/uv-vis-uv-vis-nir-spectroscopy/uv-vis-uv-vis-nir-accessories/cary-universal-measurement-accessory-uma?srsltid=AfmBOopP9IAbCEZiZkXaJVXJXH0ZiZ8OoseRKRidJpi59Sv0nmZCoMCq}}
. This spectrometer allows us to run a wavelength scan while maintaining a tightly defined geometry at a small angle (e.g., $6^\circ$). The UMA sets the sample angle precisely ($0.02^\circ$ resolution) and measures specular reflectance by placing the detector at twice the sample angle (i.e., $6^\circ$ incidence and a $12^\circ$ detector angle), which provides a repeatable “near-normal” specular reflectance spectrum.

\subsection{Measuring Performance on Micro-optics.}
\label{sec:measurement}

~~~~The Agilent Cary setup has a minimum spot size of $2,\mathrm{mm} \times 2,\mathrm{mm}$, which is much larger than the micro-lenses used in this study ($\sim 250,\mu\mathrm{m}$). Therefore, we cannot directly use the spectrometer for these measurements. Instead, we use two different methods to extract the reflectance information.

\subsubsection{Method 1: Measure coating on different substrate and then model on IPX}
\label{m1}

~~~~In this method, we place a fused silica slide in the ALD chamber along with our samples. We keep the slide and the samples in the chamber at the same time to minimize differences in the AR coating. We then measure the reflectance versus wavelength curve using the Agilent Cary system and extract the graded profile of the coating using the method described in \Cref{sims}.

Using the extracted graded profile \Cref{fig:doubles} , we simulate the reflectance and transmission versus wavelength for the microlenses by adding a base layer identical to that of the microlens in the simulation, in our case LFAST lenses can be approximated as a $250\mu m$ thick IPX Clear layer over a fused silica substrate. This yields the reflectance performance of the microlenses with the AR coating \Cref{fig:mau}.

\subsubsection{Method 2: Relative reflection measurements on microlenses}

~~~~For this method, we measure the reflected intensity, $I_R$, using the LabRAM Soleil Horiba confocal Raman microscope in micro-reflectance (micro-UV--Vis) (source is reliable up to 850nm). We illuminate the sample with a broadband source through the microscope objective, collect the reflected light from a selected region through the same objective ($30^\circ$ collection in this work), and send it to the spectrometer. The objective and confocal aperture set a $10\mu \mathrm{m}$ spot size, which allows measurements on individual micro-optic features. However, this setup does not allow us to measure the total incident light, $I_0$. Due to the highly anti reflective nature of our coating, the Raman spectrograph had to be set to $30^\circ$ to be able to perform measurements.  

Reflectance ($R$) is defined as
\begin{equation}
R = \frac{I_R}{I_0}.
\end{equation}

Because we cannot directly measure $I_0$, we instead use an identical uncoated micro-lens sample as a reference and perform the same measurement. This gives us the reflected intensity $I_x$. Using the AR-coated sample's reflected intensity $I_{AR}$ and $I_x$, As shown in \Cref{fig:mau} A we can compute the relative reflectance ($R_{x\text{-}AR}$) as,

\begin{equation}
R_{x\text{-}AR} = \frac{I_{AR}}{I_x}.
\end{equation}

The reflectance of the uncoated sample can then be calulated using  $R_x$ and the transfer matrix method\cite{GTT}, treating the uncoated sample as a simple two-layer system \Cref{fig:mau} B.

Since $R_x = \frac{I_x}{I_0}$, the absolute reflectance of the AR coating \Cref{fig:mau}, $R_{AR}$, is given by
\begin{equation}
R_{AR} = \frac{I_{AR}}{I_0}
= \frac{I_x}{I_0}\times \frac{I_{AR}}{I_x}
= R_x \times R_{x\text{-}AR}.
\end{equation}

\subsubsection{Comparing Method 1 and Method 2}

\begin{figure} [ht]
   \begin{center}
   \begin{tabular}{c} 
   \includegraphics[height=9cm]{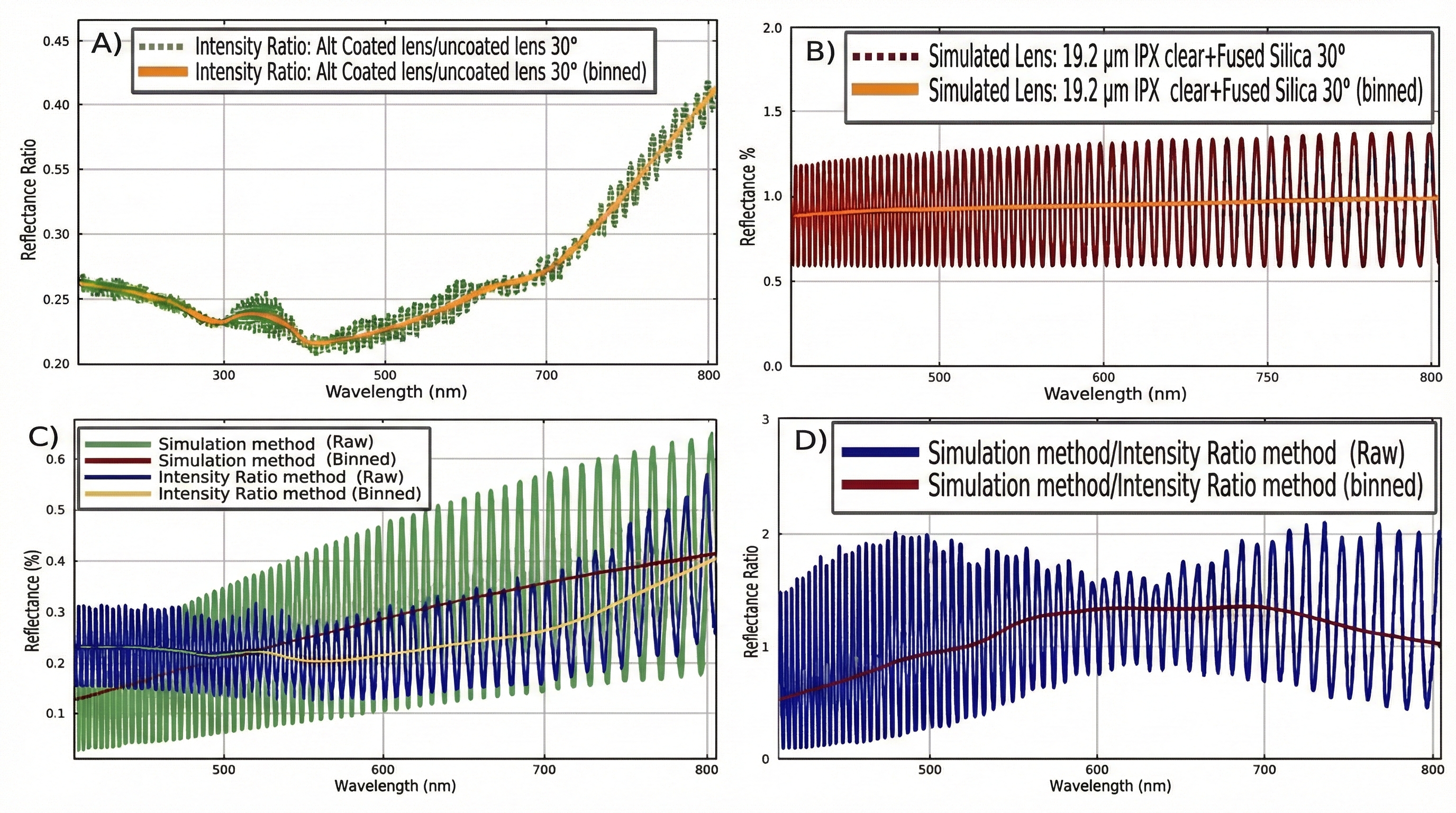}
   \end{tabular}
   \end{center}
   \caption[doubles]{\label{fig:mau} A) Intensity count ratio of AR coated lenses to Uncoated lenses (equivalent to Relative Reflectance). The Cyan graph shows the raw results and the red line shows the phase averaged result.
   
   B) Reflectance profile of a simulated Stack of $19.2\mu m$ IPX clear layer over Fused Silica ($SiO_2$) substrate. This is done to estimate the performance of the LFAST micro-lenses. The blue curve is the raw performance and the red curve is the binned performance.
   
   C) Reflectance of AR coated lenses calculated using Method 1 (Cyan and red) and Method 2 (Blue and Yellow)
   
   D) Ratio of Reflectance from Method 1 (Simulations) over Method 2 (Relative reflectance). The blue plot shows the raw data and the red plot shows the phase average results. The Ratio is within $1\pm0.5$, indicating good agreement between the two methods.}
   \end{figure}

\Cref{fig:mau} C shows the reflectance profile for the AR coated microlenses using the two methods. Both methods show that the average reflectance is within 0.6\% in the visible range, and the binned measurements are within 0.4\%, Indicating our coatings are highly anti-reflective. The average percentage deviation (binned) between the methods is 9\% \Cref{fig:mau} D. 

\section{Discussion and Conclusions}
\label{sec:conc}

Our results show high coating performance across most of the wavelength band, with a slight trade-off at shorter wavelengths. With the $SiO_2$ underlayer and Alumina Grass, the effective thickness of the AR coating can be as high as\ $\sim$350 nm. therefore, we observe scattering at shorter wavelengths, which reduces the transmission in the UV-blue region.

The LFAST application also puts a constraint on the coating geometry. We can coat only one side of the lenses because the back side will be connected to silica optical fibers. If the AR coating is on the backside, it would rather reduce the transmission because the refractive indices of the fused silica substrate and the optical fibers are nearly identical.

Despite these constraints, this approach offers broad practical advantages. We are able to produce highly conformal and uniform coatings, and we can coat and bloom many samples in parallel. In future work, since alumina grass can degrade/corrode when exposed to moisture, A top coat of a water-resistant material such as silica or titanium dioxide\cite{Aby} can be added over the AR layer to limit further corrosion from environmental exposure.

In this paper, we present an AR-coating approach for micro-optic arrays using atomic layer deposition and the anti-reflective behavior of hot-water-treated alumina. We use simulations to study how a graded-index coating and a $SiO_2$ underlayer broaden anti-reflective performance across a wider wavelength range. We also propose two measurement methods that quantify coating performance without requiring advanced nanoscale spectrometers. Together, these results provide a framework to design, model, and evaluate Alumina Grass AR coatings for micro-optic arrays under measurement and fabrication constraints. The development of such techniques will enable high efficiency bespoke and complex micro-optics for astronomical applications.
d\clearpage
\acknowledgments 

We acknowledge funding support from the Penn State Materials Research Institute for this program. The LFAST
project is supported by Schmidt Sciences LLC. The Center for Exoplanets and Habitable Worlds is supported by
the Pennsylvania State University, the Eberly College of Science. 

\bibliography{report} 
\bibliographystyle{spiebib} 

\end{document}